\documentclass[twocolumn]{aastex6}
\usepackage{verbatim}
\usepackage{aas_macros}
\usepackage{hyperref}
\usepackage{natbib}
\usepackage{graphicx}
\usepackage{epsfig}
\usepackage{amsmath}
\usepackage[english]{babel}

\begin{document}


\title{Quiescence correlates strongly with directly-measured black hole mass in central galaxies}

\author{Bryan A. Terrazas\altaffilmark{1}, Eric F. Bell\altaffilmark{1}, Bruno M. B. Henriques\altaffilmark{2}$^{,}$\altaffilmark{3}, Simon D. M. White\altaffilmark{3}, Andrea Cattaneo\altaffilmark{4}, \\Joanna Woo\altaffilmark{2}}
\shortauthors{Terrazas et al.}

\altaffiltext{1} {Department of Astronomy, University of Michigan, Ann Arbor, MI 48109, USA}
\altaffiltext{2} {Department of Physics, Institute for Astronomy, ETH Zurich, 8093 Zurich, Switzerland}
\altaffiltext{3} {Max-Planck-Institut f{\"u}r Astrophysik, D-85741 Garching, Germany}
\altaffiltext{4} {GEPI, Observatoire de Paris, CNRS, 75014 Paris, France}

\begin{abstract}
Roughly half of all stars reside in galaxies without significant ongoing star formation. However, galaxy formation models indicate that it is energetically challenging to suppress the cooling of gas and the formation of stars in galaxies that lie at the centers of their dark matter halos. In this Letter, we show that the dependence of quiescence on black hole and stellar mass is a powerful discriminant between differing models for the mechanisms that suppress star formation. Using observations of 91 star-forming and quiescent central galaxies with directly-measured black hole masses, we find that quiescent galaxies host more massive black holes than star-forming galaxies with similar stellar masses. This observational result is in qualitative agreement with models that assume that effective, more-or-less continuous AGN feedback suppresses star formation, strongly suggesting the importance of the black hole in producing quiescence in central galaxies.
\end{abstract}


%% subject header.  Objects should be in the appropriate "individual"
%% headers (e.g. quasars: individual, stars: individual, etc.) with the
%% additional provision that the total number of headers, including each
%% individual object, not exceed six.  The \objectname{} macro, and its
%% alias \object{}, is used to mark each object.  The macro takes the object
%% name as its primary argument.  This name will appear in the paper
%% and serve as the link's anchor in the electronic edition if the name
%% is recognized by the data centers.  The macro also takes an optional
%% argument in parentheses in cases where the data center identification
%% differs from what is to be printed in the paper.


\section{Introduction}
\label{sec:Intro}

Galaxy surveys have revealed the dramatic growth of the quiescent, non-star-forming galaxy population with cosmic time \citep[e.g.,][]{mms2013}. Despite the high present abundance of quiescent galaxies, the relative importance of possible physical drivers of galaxy-wide suppression of star formation remains uncertain. In a cosmological context, gas cooling and accretion into the center of a dark matter halo fuels ongoing star formation. Thus, the onset of quiescence means that gas is somehow removed from the galaxy and that gas cooling is offset by some source of heat. Unlike satellites, galaxies in the center of a halo's potential well -- hereafter referred to as central galaxies -- must eject and heat their gas without relying on interactions with the hot, diffuse medium present in other halos, groups, and clusters \citep{tlb2013}. This implies stringent energetic requirements not easily met by stellar feedback \citep[e.g.][]{bbm2006}.

Heating mechanisms proposed for central galaxies include ejected gas from supernovae Ia (SNIa) and stellar winds \citep[e.g.][]{hqm2012}, virial shock heating \citep[e.g.][]{bdn2007}, gravitational heating \citep[e.g.][]{jno2009}, and -- currently the most popular explanation -- feedback from active galactic nuclei \citep[AGN,][]{kh2000,dsh2005, csw2006, cfb2009, f2012}.

One powerful approach towards characterizing the importance of different physical drivers of quiescence in central galaxies is to measure the correlation between quiescence and a range of galaxy properties that could affect the balance between heating and cooling. For example, cooling and gas accretion depend strongly on halo mass, and would thus be expected to correlate with stellar mass (with significant scatter; see \citealp{tbh2016}). Heating or gas ejection could correlate with a variety of properties: halo mass due to virial shock heating or gravitational quenching, stellar mass due to SNIa and stellar feedback, or black hole mass due to AGN feedback.

With these concerns in mind, many studies have explored how quiescence correlates with a variety of quantities: for example, stellar mass, halo mass, surface density, inferred velocity dispersion, \citet{s1963} index, and bulge mass \citep{khw2003, fvf2008, bvp2012, lws2014, bme2014, wdf2015, mwz2016}. The latter quantities are expected to correlate with the prominence of a supermassive black hole \citep{kh2013}, in support of the idea that AGN feedback is an important driver of quiescence. Yet, correlating quiescence with directly-measured black hole mass would be a clearer and more critical test of AGN feedback. With the number of dynamical black hole mass measurements increasing each year, such an exercise has now become possible.

The goal of this Letter is to characterize the physical drivers of quiescence by studying the observed distribution of star-forming and quiescent central galaxies as a function of their central black hole mass and stellar mass (\S\ref{sec:obsdata}) and comparing those findings with the results from four galaxy formation models (\citealp{hwt2015}, \S\ref{sec:henriques}; Illustris -- \citealp{vgs2014}, \S\ref{sec:illustris}; EAGLE -- \citealp{scb2015}, \S\ref{sec:EAGLE}; and GalICS -- \citealp{cdd2006}, \S\ref{sec:GALICS}). We then describe (\S\ref{sec:results}) and discuss (\S\ref{sec:disc}) the apparent agreement between observations and models that use effective, more-or-less continuous AGN feedback to halt star formation. We assume the standard cosmology in order to be consistent with our compiled observational distances: $\Omega_{M}$ = 0.3, $\Omega_{\Lambda}$ = 0.7, and H$_{0}$ = 70 km/s/Mpc.

\section{Data}
\label{sec:data}

\subsection{Observational estimates of black hole masses, stellar masses and star formation rates}
\label{sec:obsdata}

Dynamical estimates of black hole masses ($M_{\rm{BH}}$) are heterogeneous, coming from stellar dynamics, gas dynamics, masers, and reverberation mapping techniques. We adopt the $M_{\rm{BH}}$ estimates compiled by \citet{soe2016}, supplemented by \citet[and references therein]{v2016}. Our conclusions are insensitive to the particular compilation that we adopt. We select central galaxies by identifying the brightest or only members of their group within a $\sim$1 Mpc radius in order to omit the effects of quenching unique to satellites. Finally, we choose nearby galaxies within $\sim$150 Mpc (z $\lesssim$ 0.034). Our final sample includes 91 central galaxies.

Stellar masses ($M_{*}$) were estimated using extinction-corrected `total' $K_{\rm{s}}$ apparent magnitudes from the 2MASS Redshift Survey \citep{hmm2012}. We adopt a single K-band stellar $M_{*}/L_{\rm{K}}$ ratio of 0.75, the average value for the luminous galaxies studied by \citet{bmk2003}, adjusted to a \citet{c2003} IMF. The variation in $M_{*}/L_{\rm{K}}$ is expected to be too small to significantly affect our results \citep{bmk2003}.

The chief observational novelty of our analysis is the use of star formation rates (SFRs) to characterize quiescence in conjunction with directly-detected black hole masses. We calculate far-infrared (FIR) derived SFRs using IRAS \citep[][see also corrections to \citealp{kgk1989} in NED by Knapp 1994]{rls1988, m1990, ssm2004, sh2009}. As discussed in \citet{b2003}, FIR-derived SFRs are most appropriate for relatively massive galaxies with significant dust contents, since ultraviolet (UV) or H$\alpha$ fluxes are typically strongly attenuated by dust. The FIR is also less susceptible to contamination from AGN than mid-IR or radio SFR estimates. Equation A1 in \citet{b2003} uses 60 and 100 $\mu$m fluxes to estimate the FIR flux. Non-detections are estimated using the ratios $f_{60}/f_{70}$ = 0.88, $f_{60}/f_{100}$ = 0.39, $f_{60}/f_{25}$ = 7.19, $f_{60}/f_{12}$ = 11.0, which are derived from a large number of local galaxies. The 70 $\mu$m measurements are from Spitzer/MIPS \citep{tbm2009, dcj2009}. We then estimate the total infrared (TIR) flux via TIR = 2 $\times$ FIR \citep{b2003}. The TIR-derived SFR is calculated using Equation 12 in \citet{ke2012},
\begin{equation}
\label{eq:SFR}
\rm{log}_{10} \rm{SFR}_{\rm{TIR}}\: (M_{\odot} \: \rm{yr}^{-1}) = \rm{log}_{10} \textit{L}_{\rm{TIR}} - 43.41
\end{equation}
where L$_{\rm{TIR}}$ is the TIR luminosity calculated using our TIR flux estimates and the distances to the galaxies. Galaxies with no infrared detections or detections that result in SFR/$M_{*}$ $<$ 10$^{-13}$ yr$^{-1}$ are taken as upper limits. We adopt a factor of two uncertainty for our SFR values \citep{b2003}. We have confirmed that hybrid TIR+UV SFRs for those galaxies that have measured UV fluxes yield similar results to TIR-only SFRs.

%%%%%%%%%%%%%%%%%%%%%%%%%%%%%%%%%%%%%%%%%%%%%%%%%%%%%%
%4 Model Comparison Plot
\begin{figure*}
\epsfxsize=17.5cm
\epsfbox{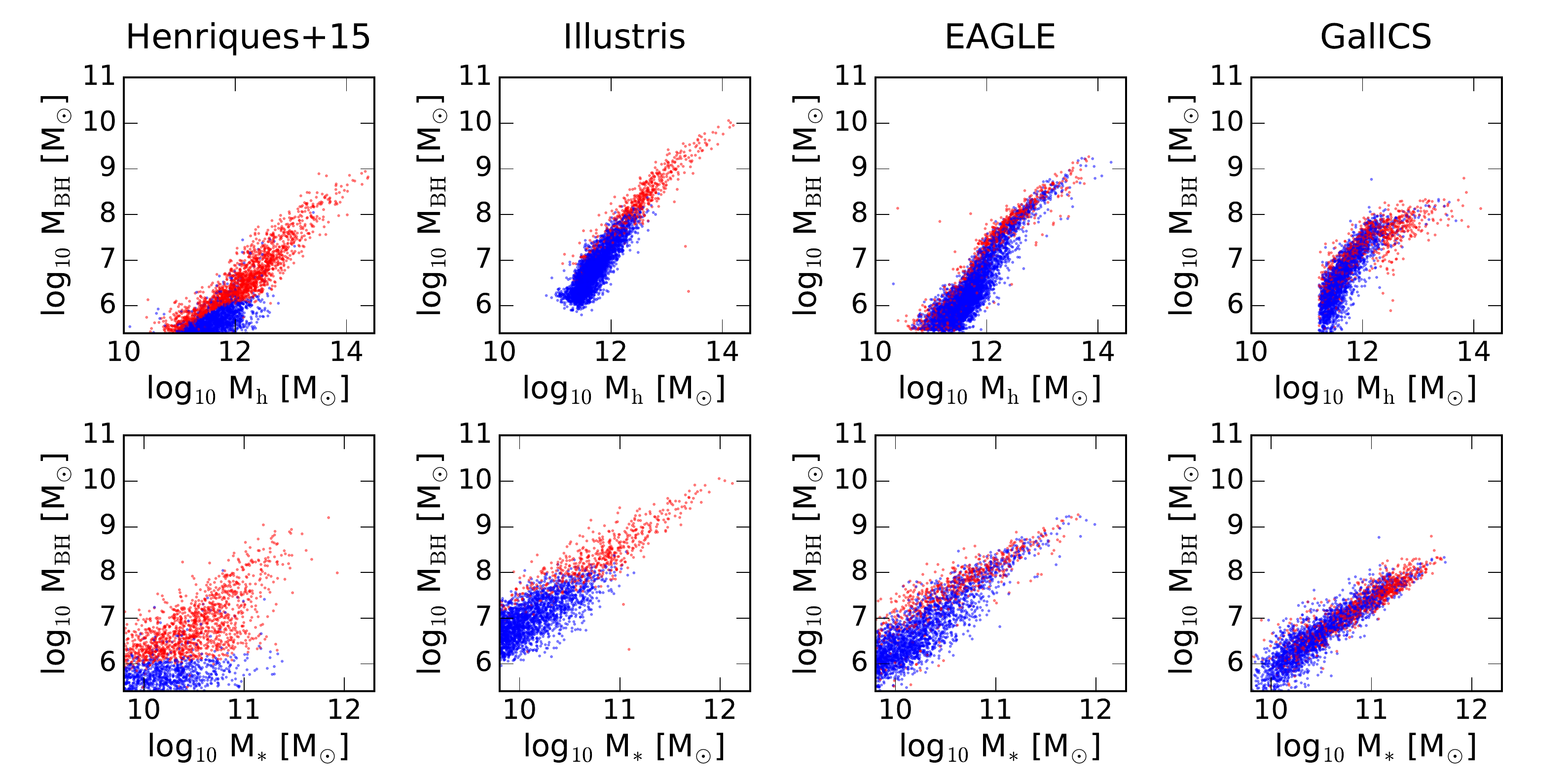}
\caption{$M_{\rm{BH}}$ as a function of $M_{\rm{h}}$ (upper panels) and $M_{*}$ (lower panels) for the \citet{hwt2015}, Illustris, EAGLE, and GalICS models. Blue and red points indicate star-forming and quiescent galaxies, respectively, chosen via the SFR selection described in \S\ref{sec:data}.}
\label{fig:4modelplot}
\end{figure*}
%%%%%%%%%%%%%%%%%%%%%%%%%%%%%%%%%%%%%%%%%%%%%%%%%%%%%%

\subsection{The Henriques et al. (2015) Semi-Analytic Model}
\label{sec:henriques}

\citet{hwt2015} developed a semi-analytic model that uses the Millennium Simulations \citep{swj2005, bsw2009} to provide the dark matter framework in which they embed their analytic prescriptions for the evolution of gas and stars. Quiescence in the \citet{hwt2015} model is primarily a result of heating from continuous radio-mode AGN feedback, which halts the cooling of the circumgalactic medium onto the galaxy's disk. This effectively cuts off the fuel needed to form stars. Analytically, the model is built so that the balance between heating and cooling depends strongly on $M_{\rm{BH}}$ and only somewhat on the hot gas mass, which correlates strongly with halo mass \citep[$M_{\rm{h}}$; see Figure 1 in][]{tbh2016}.

\subsection{The Illustris Hydrodynamic Simulation}
\label{sec:illustris}

The Illustris Project is a series of large-scale hydrodynamic simulations of galaxy formation \citep{vgs2014}. These simulations use the moving-mesh technique {\small AREPO} \citep{s2010} to follow individual particles in order to model the baryonic physics relevant to galaxy evolution. Similarly to the \citet{hwt2015} model, galaxies in Illustris depend on a balance between heating and cooling in order to determine quiescence. Radio-mode AGN feedback transfers heat to the atmospheres around galaxies via the expansion of hot bubbles emanating from the black hole. The amount of thermal energy transferred depends on the growth of $M_{\rm{BH}}$ in the radio mode \citep{svg2015}.

\subsection{The EAGLE Hydrodynamic Simulation}
\label{sec:EAGLE}

The EAGLE Project \citep{scb2015} is a suite of hydrodynamic simulations that use a modified version of the smoothed particle hydrodynamics code {\small GADGET 3} \citep{s2005} to model the physics of galaxy formation. They include one mode of AGN feedback most closely resembling the quasar mode, which the model depends upon to suppress star formation in high mass galaxies. Thermal energy is injected at a rate proportional to the gas accretion rate, which depends on $M_{\rm{BH}}$ along with the properties of the gas around it. In this model, AGN feedback works stochastically through short-lived events that inject heat into the interstellar medium of the galaxy.

\subsection{The GalICS Semi-Analytic Model}
\label{sec:GALICS}

We use the implementation of the GalICS semi-analytic model described in \citet{cdd2006}. In this model, star formation is shut off above a critical halo mass, $M_{\rm{h,crit}}$ $\sim$ 10$^{12}$ M$_{\odot}$, which represents the sharp transition from free-falling cold-mode to shock-driven hot-mode gas accretion onto the galaxy. At larger halo masses, cold gas in the galaxy is heated to the virial temperature and added to the hot gas component. Once shock-heated gas is available, AGN are able to provide a source of feedback through inefficient accretion and maintain the high temperatures of the gas in order to prevent cooling and subsequent star formation.

\vspace{7mm}
In order to provide a common method for differentiating star-forming and quiescent galaxies at $z \sim 0$ for all models, we identify a best fit line to the star-forming main sequence for all four models and observations of a representative sample of local galaxies without $M_{\rm{BH}}$ measurements. We define quiescent galaxies as those that lie a factor of 4 or more below this line.

%%%%%%%%%%%%%%%%%%%%%%%%%%%%%%%%%%%%%%%%%%%%%%%%%%%%%%
%Observational plot
\begin{figure*}
\begin{center}
\epsfxsize=11cm
\epsfbox{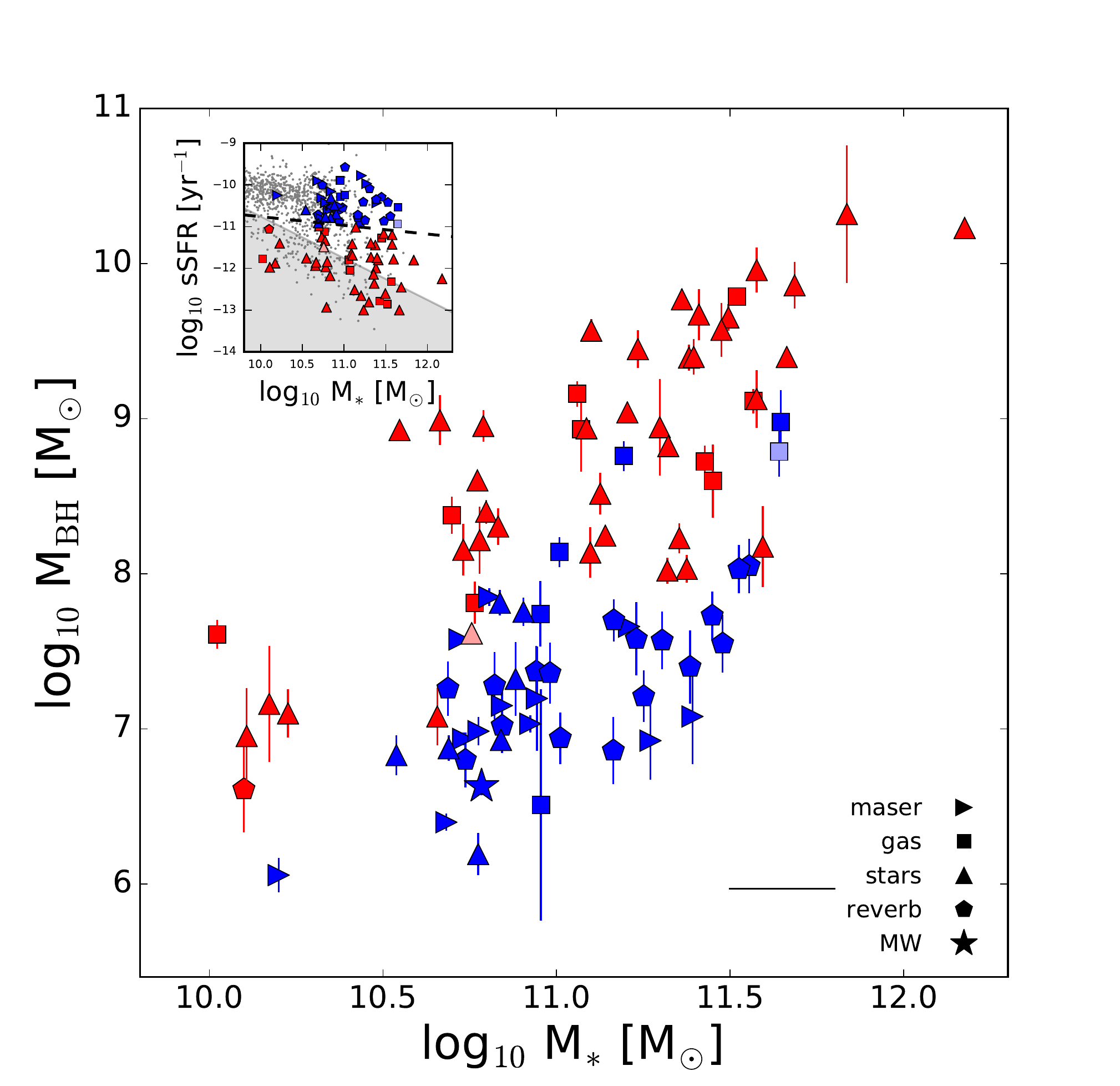}
\caption{Directly-measured $M_{\rm{BH}}$ as a function of $M_{*}$ for star-forming (blue) and quiescent (red) central galaxies in the nearby universe (z $<$ 0.034). The black line indicates the uncertainty on $M_{*}$. The inset plot shows the sSFR-$M_{*}$ plane for a selection of local galaxies (gray points) and for all galaxies in our sample (colored points). The shaded region indicates where the selection of local galaxies is no longer complete. Lighter colored points represent mid-IR-derived SFRs that should be taken as upper limits.}
\label{fig:obsresult}
\end{center}
\end{figure*}
%%%%%%%%%%%%%%%%%%%%%%%%%%%%%%%%%%%%%%%%%%%%%%%%%%%%%%

\section{Results}
\label{sec:results}

Many observed galaxy properties correlate with each other and the mechanisms behind quiescence may be complex. Accordingly, we first use the models to generate intuition about how the physical drivers of quiescence may impact observational correlations before examining the observations.

\subsection{A comparison between models}
\label{sec:results_models}

In Fig.~\ref{fig:4modelplot}, we show the physically-important but currently unobservable $M_{\rm{BH}}$--$M_{\rm{h}}$ plane in the upper panels, and the observable $M_{\rm{BH}}$--$M_{*}$ plane in the lower panels for all models. We find a variety of distributions in $M_{\rm{BH}}$--$M_{*}$--$M_{\rm{h}}$ parameter space. 

The \textit{quantitative} differences in normalization result from the calibration of the $M_{\rm{BH}}$ growth efficiencies to different $M_{\rm{BH}}$-galaxy relations. The feedback efficiencies that regulate star formation are largely decoupled from the $M_{\rm{BH}}$ growth efficiencies in all models. This suggests that differences in the calibration of the $M_{\rm{BH}}$ growth efficiencies would not affect which galaxies are star-forming or quiescent. Therefore, the crucial features for our purposes are \textit{qualitative} differences in the distribution of star-forming and quiescent galaxies between models, which can be used as a diagnostic of the physical drivers of quiescence in these models.

We find that the \citet{hwt2015} and Illustris models show a qualitatively similar divide between star-forming and quiescent galaxies, a division that depends strongly on $M_{\rm{BH}}$ and much less strongly on $M_{\rm{h}}$ and $M_{*}$. In these models, a quiescent galaxy almost always has a larger black hole than a star-forming galaxy due to the connection between the $M_{\rm{BH}}$ and the heating rate from long-lived radio-mode AGN feedback. While the \citet{hwt2015} model demonstrates this behavior by construction \citep{tbh2016}, this result emerges from Illustris quite naturally from their hydrodynamic recipes where there is no explicit link between the heating rate and galaxy properties such as $M_{\rm{BH}}$ or $M_{\rm{h}}$.

The EAGLE Simulation shows similar behavior where quiescent galaxies are more likely to have massive black holes. Star-forming galaxies, however, span the entire range of $M_{\rm{BH}}$ and $M_{*}$, where galaxies with massive black holes can still be star-forming. This is confirmed in studies of the EAGLE Simulation showing that the passive fraction at higher $M_{*}$ is too low compared to observations \citep{fbt2015}. We posit that the short-lived nature of the feedback that heats the interstellar medium in their model does not stop gas cooling in between these events, where star formation can continue in galaxies with a non-accreting yet massive black hole \citep[see also][]{ttb2016}.

Finally, the GalICS model shows overlapping distributions of star-forming and quiescent galaxies on the $M_{\rm{BH}}$--$M_{*}$ plane with quiescent galaxies preferentially at higher $M_{*}$. The quenching mechanism is evident in the $M_{\rm{BH}}$--$M_{\rm{h}}$ plane where there is a dramatic deficit of star-forming galaxies above $M_{\rm{h}} \sim$ 10$^{12.3}$ M$_{\odot}$. The assumption of a critical $M_{\rm{h}}$ at which star formation stops results in $M_{\rm{BH}}$ having little to no importance for quiescence in this model.

%%%%%%%%%%%%%%%%%%%%%%%%%%%%%%%%%%%%%%%%%%%%%%%%%%%%%%
%sigma/Mbulge plots
\begin{figure*}
\begin{center}
\epsfxsize=11.5cm
\epsfbox{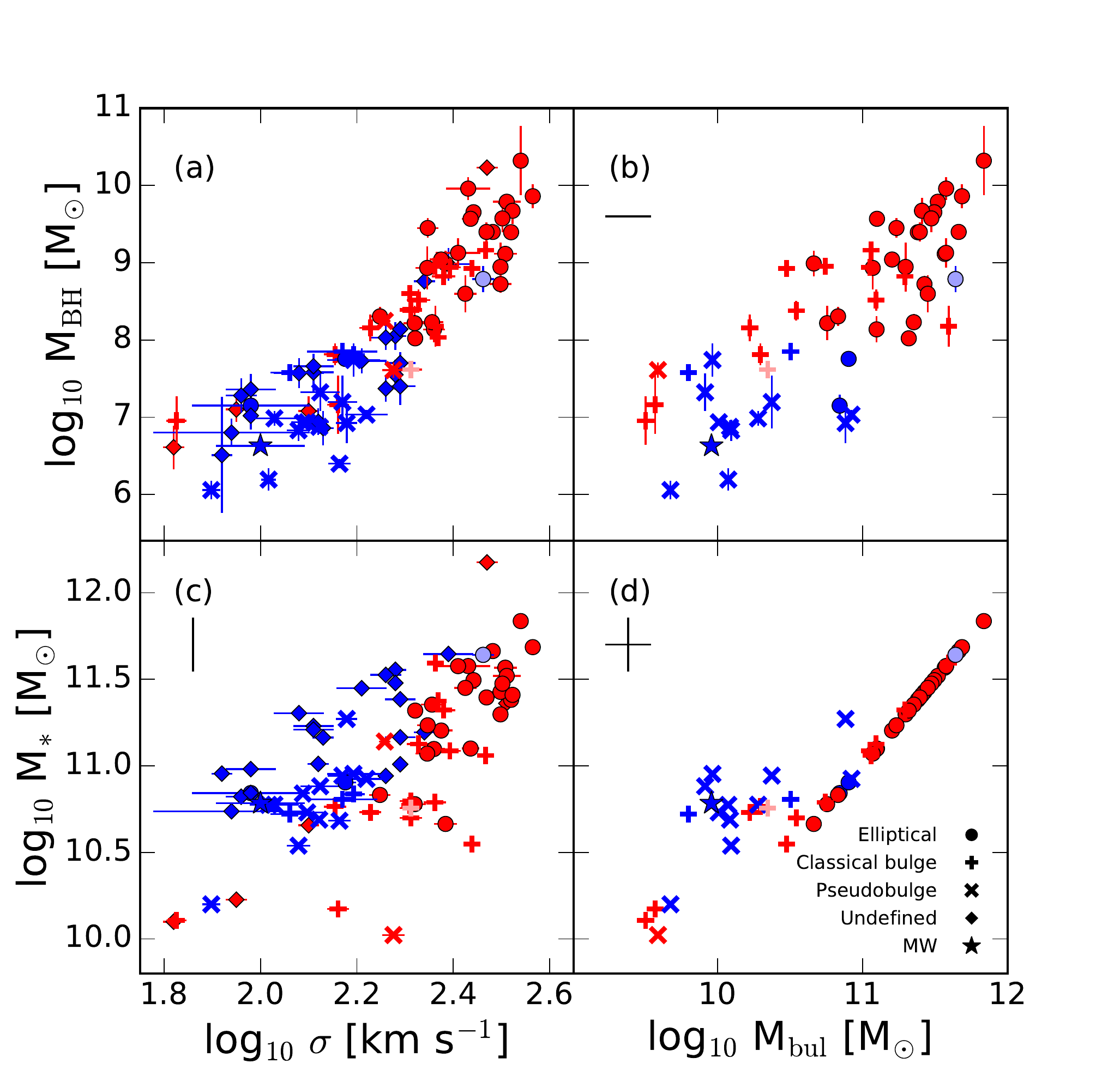}
\caption{A collection of panels showing the (a) $M_{\rm{BH}}$-$\sigma$, (b) $M_{\rm{BH}}$-$M_{\rm{bul}}$, (c) $M_{*}$-$\sigma$, and (d) $M_{*}$-$M_{\rm{bul}}$ relations for star-forming (blue) and quiescent (red) galaxies, where we omit those with no $\sigma$ or $M_{\rm{bul}}$ measurements. The black lines indicate the uncertainties on $M_{*}$ and $M_{\rm{bul}}$. Morphologies, if defined, are from \citet{soe2016}. Lighter colored points represent mid-IR-derived SFRs that should be taken as upper limits.}
\label{fig:bulgesigma}
\end{center}
\end{figure*}
%%%%%%%%%%%%%%%%%%%%%%%%%%%%%%%%%%%%%%%%%%%%%%%%%%%%%%

\subsection{Observational evidence of the link between black hole mass and quiescence}
\label{sec:results_obs}

Given the diagnostic power of the lower panels of Fig.~\ref{fig:4modelplot}, we present a direct observational counterpart in Fig.~\ref{fig:obsresult}. The inset plot shows the criterion (black dashed line) we choose in \S\ref{sec:data} for identifying star-forming (blue) and quiescent (red) galaxies when plotting the specific star formation rate (SFR/$M_{*}$, sSFR) against the $M_{*}$ while also showing a subset of local galaxies without directly-measured black hole masses as gray points. The shaded region represents where the subset of local galaxies is no longer complete due to the detection limit of the infrared measurements. 

Fig.~\ref{fig:obsresult} shows a pronounced divide between star-forming and quiescent galaxies where quiescent galaxies have more massive black holes than their star-forming counterparts. In addition, there is a $M_{*}$ dependence to this divide, where lower $M_{*}$ galaxies can be quiescent at lower $M_{\rm{BH}}$ than higher $M_{*}$ galaxies.

Comparing our observational result with the model data in the lower panels of Fig.~\ref{fig:4modelplot}, we find that real galaxies more closely resemble models in which effective, more-or-less continuous AGN feedback quenches star formation in central galaxies -- namely, the \citet{hwt2015} model (\S\ref{sec:henriques}) and Illustris (\S\ref{sec:illustris}). As we have described in \S\ref{sec:results_models}, these models result in a pronounced divide between star-forming and quiescent galaxies with little scatter -- similar to Fig.~\ref{fig:obsresult}. We note that the EAGLE Simulation produces a more similar $M_{*}$ dependence with regards to the divide, yet fails to replicate the separation between star-forming and quiescent galaxies on this plane.

\subsection{Bulge Mass and Velocity Dispersion}
\label{sec:bulgesigma}

The motivation for exploring the relationship between $M_{*}$, $M_{\rm{BH}}$, and quiescence was to test the importance of $M_{\rm{BH}}$ in driving quiescence. Previous works have linked quiescence with quantities that correlate with $M_{\rm{BH}}$, such as velocity dispersion \citep[$\sigma$, e.g.,][]{fvf2008} or bulge mass \citep[$M_{\rm{bul}}$, e.g.,][]{bep2014}. As such, whether $\sigma$ or $M_{\rm{bul}}$ correlates better with quiescence than $M_{\rm{BH}}$ may provide important physical insight. 

This question is explored in Fig.~\ref{fig:bulgesigma}, where we present the $M_{\rm{BH}}$--$\sigma$, $M_{\rm{BH}}$--$M_{\rm{bul}}$, $M_{*}$--$\sigma$, and $M_{*}$--$M_{\rm{bul}}$ relations for our sample, omitting those with no $\sigma$ or $M_{\rm{bul}}$ measurements. $\sigma$ was provided by \citet{v2016} and $M_{\rm{bul}}$ was obtained by adopting the bulge-to-total ratios in $K_{\rm{s}}$ band found in \citet{kh2013}. Morphologies, if defined, are from \citet{soe2016} and are indicated using different symbols.

Figure~\ref{fig:bulgesigma}c shows that quiescence correlates well with $\sigma$ at a given $M_{*}$. This correlation is as strong as the correlation between $M_{\rm{BH}}$ and quiescence, possibly due to the tight correlation between $\sigma$ and $M_{\rm{BH}}$ (Figure~\ref{fig:bulgesigma}a). However, $\sigma$ may also directly influence the ability of galaxies to form stars. \citet{mbt2009} found that shear modulates star formation efficiency by factors of a few in highly concentrated galaxies. Yet, in the context of cosmological models, this effect is insufficient to drive quiescence, instead requiring a much larger input of energy -- generally from AGN feedback -- to keep cold gas out of galaxies. Further study of cold gas supply and SFRs as a function of $M_{\rm{BH}}$ and $\sigma$ may help illuminate the relationship between these two factors and quiescence.

Figure~\ref{fig:bulgesigma}b/d shows that $M_{\rm{bul}}$ correlates poorly with quiescence for our sample. Since $M_{\rm BH}$ correlates slightly better with $M_{\rm bul}$ than with $M_*$, one may expect that higher $M_{\rm BH}$ in quiescent galaxies are a result of larger bulge-to-total ratios. Figure~\ref{fig:bulgesigma}b shows that this is not entirely the case -- $M_{\rm BH}$ is higher in quiescent galaxies even at fixed $M_{\rm bul}$.

Furthermore, we find that quiescence is common in elliptical galaxies and galaxies with classical bulges, whereas star-forming galaxies tend to have pseudobulges. This may suggest that the processes leading to the growth of classical bulges (e.g., mergers, misaligned gas infall) may result in more effective $M_{\rm BH}$ growth than those that create pseudobulges.

\section{Discussion}
\label{sec:disc}

The goal of this Letter is to probe the physical drivers of quiescence by looking for correlations between $M_{*}$, $M_{\rm{BH}}$, and sSFR. Our main observational result is that central quiescent galaxies contain more massive black holes than their star-forming counterparts, with the boundary between these groups also having a dependence on $M_{*}$. When comparing our results with four galaxy formation models, we find the best agreement with models that simulate more effective and long-lived AGN feedback. Taken together, this analysis suggests that the central black hole has an essential role in shutting off star formation.

The clear division in the $M_{\rm{BH}}$--$M_{*}$ plane between star-forming and quiescent central galaxies is a powerful test of prescriptions for gas cooling, gas heating, and quiescence in models. Our results suggest that models that do not suppress star formation via quasi-continuous black hole-driven feedback will not produce a strong enough correlation between quiescence and $M_{\rm{BH}}$.

This work connects well with previous studies by exploring much more explicitly the interplay between SFR and $M_{\rm{BH}}$. In \citet{rv2015} and \citet{sgm2015}, the morphology of galaxies was shown in the $M_{\rm{BH}}$--$M_{*}$ plane. In both works, early and late type galaxies inhabit clearly distinct parts of the $M_{\rm{BH}}$--$M_{*}$ plane. Our work is consistent with theirs, and frames the interpretation of this behavior much more explicitly in terms of a dominant role for AGN feedback in driving quiescence. 

Other studies have used indirect proxies for $M_{\rm{BH}}$. For example, \citet{bep2014} used indirect estimates of $M_{\rm{BH}}$ from $\sigma$ and $M_{\rm{bul}}$ for central galaxies to find a transition between mostly active to mostly passive galaxies within $\sim$1.5 orders of magnitude of $M_{\rm{BH}}$. This transition appears broader than ours, which may be influenced by uncertainties in their $M_{\rm{BH}}$ estimates, and may indicate, as our results seem to, that quiescence is a function of multiple parameters such as both $M_{\rm{BH}}$ and $M_{*}$.

Our sample is selected to have dynamically-derived $M_{\rm{BH}}$ estimates and includes both inactive galaxies (favoring larger $M_{\rm{BH}}$ to maximize detectability) and active galaxies (probing lower $M_{\rm{BH}}$ systems that are accreting gas and preferentially located in star-forming galaxies). Sample selection is currently very heterogeneous, making it impractical at this stage to impose observationally-motivated selections on our model samples (e.g., to make mock observations for Fig.~\ref{fig:4modelplot}). As observational methods improve and more representative measurements become available over a wider range of galaxy types, it will be important to check if this apparent division between star-forming and quiescent galaxies in the $M_{\rm{BH}}$--$M_{*}$ plane remains.

\section{Conclusions}
\label{sec:conc}

Cosmological models of galaxy formation predict that the relationship between quiescence, $M_{\rm{BH}}$, and $M_{*}$ is a crucial discriminator between models and a sensitive probe of the drivers of quiescence. We compare directly-measured $M_{\rm{BH}}$, $M_{*}$, and other properties of a sample of star-forming and quiescent galaxies, finding that observed quiescent galaxies have higher $M_{\rm{BH}}$ than star-forming galaxies with similar $M_{*}$. These trends are in good qualitative agreement with models in which star formation is suppressed due to quasi-continuous heating from AGN feedback. We assert that models that do not replicate this behavior are missing an essential element in their physical recipes. Our study suggests that the central black hole is critical to the process by which star formation is terminated within central galaxies, giving credence to the AGN quenching paradigm.

\acknowledgements

B.A.T. is supported by the National Science Foundation Graduate Research Fellowship under Grant No. DGE1256260. We acknowledge helpful discussions with E. Gallo, J. Bregman, K. Gultekin, H. W. Rix, S. Faber, S. Ellison, A. Pontzen, and L. Sales. This work used the SIMBAD database at CDS, and the NASA/IPAC Extragalactic Database (NED), operated by the Jet Propulsion Laboratory, California Institute of Technology, under contract with NASA.

\end{document}